\documentclass{aa}  

\usepackage{comment}
\usepackage{graphicx}
\usepackage[dvipsnames]{xcolor}
\usepackage{subcaption}
\graphicspath{ {./figures/} }

\usepackage{txfonts}
\usepackage[%
  breaklinks = true,
  colorlinks = true,
  urlcolor   = blue,
  citecolor  = blue,
  linkcolor  = blue,
]{hyperref}
\usepackage{soul}
\usepackage{nameref}
\usepackage{makecell}
\usepackage{booktabs}
\usepackage[flushleft]{threeparttable}
\usepackage[labelfont=bf]{caption}
\usepackage{hhline}

\newcommand{\yokshi}{YS-94}
\newcommand{\syntelis}{Syntelis-19}
\newcommand{\hdiff}{Gudiksen-11}

\defcitealias{2023A&A...675A..97F}{F2023}

\begin{document}

   \title{A comparative study of resistivity models for simulations of magnetic reconnection in the solar atmosphere.\\ II. Plasmoid formation}
   \titlerunning{A Comparative Study of Resistivity Models... II. Plasmoid formation}
   \authorrunning{Ø.H.Færder et al.}

   \author{Ø. H. Færder\inst{1,2},
          D. Nóbrega-Siverio\inst{3,4,1,2}
          \and
          M. Carlsson\inst{1,2}
          }

   \institute{Rosseland Centre for Solar Physics, University of Oslo,
              PO Box 1029, Blindern, NO-0315 Oslo, Norway\\
              \email{o.h.farder@astro.uio.no}
         \and
              Institute of Theoretical Astrophysics, University of Oslo,
              PO Box 1029, Blindern, NO-0315 Oslo, Norway
         \and
             Instituto de Astrof\'isica de Canarias,    E-38205 La Laguna,  Tenerife, Spain
        \and
            Universidad de La Laguna, Dept. Astrof\'isica, E-38206 La  Laguna, Tenerife, Spain
             }

 
  \abstract
   {Plasmoid-mediated reconnection plays a fundamental role in different solar atmospheric phenomena.
   Numerical reproduction of this process is therefore essential for developing robust solar models.}
   {Our goal is to assess plasmoid-mediated reconnection across various numerical resistivity models in order to investigate how plasmoid numbers and reconnection rates depend on the Lundquist number.}
  {We used the Bifrost code to drive magnetic reconnection in a 2D coronal fan-spine topology, carrying out a parametric study of several experiments with different numerical resolution and resistivity models. We employed three anomalous resistivity models: (1) the original hyper-diffusion from Bifrost, (2) a resistivity proportional to current density, and (3) a resistivity quadratically proportional to electron drift velocity. For comparisons, experiments with uniform resistivity were also run.}
   {Plasmoid-mediated reconnection is obtained in most of the experiments. With uniform resistivity, increasing the resolution reveals higher plasmoid frequency with weaker scaling to the Lundquist number, obtaining 7.9-12 plasmoids per minute for $S_\mathrm{L}\in[1.8 \times 10^4, 2.6\times 10^5]$ with a scaling of $S_\mathrm{L}^{0.210}$ in the highest-resolution resistivity cases, transcending into Petschek reconnection in the high-$S_\mathrm{L}$ limit (where the diffusive effects of the resistivity become small compared to the non-uniform viscosity)
   and Sweet-Parker reconnection in the low-$S_\mathrm{L}$ limit. Anomalous resistivity leads to similar results even with lower resolution. The drift-velocity-dependent resistivity excellently reproduces Petschek reconnection for any Lundquist number, and similar results are seen with resistivity 
proportional to  current-density though with slightly lower reconnection rates and plasmoid numbers. 
   Among the different resistivity models applied on the given numerical resolution, the hyper-diffusion model reproduced plasmoid characteristics in closest resemblance to those obtained with uniform resistivity at a significantly higher resolution.
    }
  {}


   \maketitle
   
%

\section{Introduction}

Magnetic reconnection is a promising candidate as a mechanism for heating up the solar corona \citep[e.g.][]{1973SoPh...32...81V, 1984A&A...137...63H, 1988ApJ...330..474P}. In addition, this process has been shown to unleash some of the important phenomena in the solar atmosphere that have been successfully modelled in numerical experiments; these include Ellerman bombs (EBs) and ultraviolet (UV) bursts \citep[e.g.][]{2017ApJ...839...22H, 
    2019A&A...626A..33H,
    2017A&A...601A.122D,
    2017ApJ...850..153N,
    2019A&A...628A...8P,
    2021A&A...646A..88N},
surges and coronal jets \citep[e.g.][]{ 
    1995Natur.375...42Y, 1996PASJ...48..353Y,
    2016ApJ...822...18N, 
    2016ApJ...827....4W,
    2017Natur.544..452W,
    2022ApJ...935L..21N},
as well as flares \citep[e.g.][]{
    2001ApJ...549.1160Y,
    2023ApJ...955..105R}.

This fundamental mechanism can either be modelled as steady reconnection or non-steady, plasmoid-mediated reconnection. In the former case, one may analytically predict how the reconnection rate, among other quantities, depends on the Lundquist number $S_\mathrm{L} \equiv L v_{\mathrm{Ai}} / \eta$, where $L$ is the length of the current sheet, $v_{\mathrm{Ai}}$ the inflow Alfvén speed, and $\eta$ the resistivity of the medium. In the slow-reconnection model developed by \citet{1958IAUS....6..123S,1958NCim....8S.188S} and \citet{1963ApJS....8..177P}, where a uniform diffusion layer is assumed to cover the entire current sheet, the reconnection rate is predicted to be equal to $S_\mathrm{L}^{-1/2}$. In the fast reconnection model by \citet{1964NASSP..50..425P}, which assumes a Sweet-Parker diffusion layer that covers only a limited segment of the current sheet, the reconnection rate is found to be roughly equal to $\pi/(8\ln S_\mathrm{L})$.

Non-steady reconnection is characterised by resistive tearing instability \citep[see][]{1963PhFl....6..459F}, where magnetic islands, or plasmoids, appear rapidly along the current sheet. Plasmoid instability occurs when $S_\mathrm{L}>10^4$ \citep{2007PhPl...14j0703L}, where the current sheet gets intrinsically unstable when its inverse aspect ratio $a/L$ ---where $a$ is the current-sheet width--- passes
below a threshold value of $S_\mathrm{L}^{-1/3}$ \citep{2014ApJ...780L..19P},
which for coronal Lundquist numbers is significantly higher than the Sweet-Parker inverse aspect ratio of $S_\mathrm{L}^{-1/2}$. Therefore, Sweet-Parker reconnection is not expected to occur commonly in the upper solar atmosphere, given that any current sheet becomes unstable long before obtaining a Sweet-Parker-like aspect ratio. The Sweet-Parker reconnection rate, given a coronal Lundquist number, is also far too slow to reproduce any flare \citep[see][and references therein]{2014masu.book.....P}. Petschek-like reconnection rates have, on the other hand,  been successfully  reproduced numerically when applying a local enhancement of the resistivity in the current sheet \citep{1994ApJ...436L.197Y} or a very low, uniform resistivity \citep{2009PhPl...16a2102B}, even in the case of non-steady reconnection.

For plasmoid-mediated reconnection in an adiabatic medium, the number of plasmoids has been analytically predicted to scale with the Lundquist number as $S_\mathrm{L}^{0.375}$ \citep{2007PhPl...14j0703L}. For the non-adiabatic case, \citet{2022A&A...666A..28S} numerically found the maximum plasmoid number in a 2D Harris current sheet to scale as $S_\mathrm{L}^{0.223}$. In both cases, the number of plasmoids increases slowly with the Lundquist number. Plasmoids can therefore be expected to be quite numerous in coronal current sheets due to the relatively high Lundquist number. The presence of plasmoids in EBs, UV bursts, surges, and coronal jets has been shown both observationally \citep[e.g.,][]{2017ApJ...851L...6R, 2023A&A...673A..11R, 2019ApJ...885L..15K} and numerically \citep{2017ApJ...841...27N, 2017ApJ...850..153N, 2019A&A...626A..33H, 2019A&A...628A...8P, 2020ApJ...901..148G, 2022A&A...665A.116N, 2023RAA....23c5006L}. Numerical studies of plasmoid-mediated reconnection are therefore key to understanding  any reconnection event that may occur in the solar atmosphere.

In our previous paper \citep[hereafter \citetalias{2023A&A...675A..97F}]{2023A&A...675A..97F}, we compared three different anomalous resistivity models by applying them on a 2D magnetohydrodynamics (MHD) simulation with flux cancellation. There, we found that the models were all capable of reproducing roughly the same large-scale results in terms of current-sheet length and Poynting influx. In the present paper, we analyse the details of the plasmoid instability of these resistivity models during magnetic reconnection at the null-point of a 2D fan-spine topology and compare the results to cases with uniform resistivity. To this end, we perform a parametric study, employing different resistivity magnitudes and resolutions.
The structure of the paper is as follows. Section \ref{sec:model} describes the code and model equations used for our simulations, the different resistivity models, and the setup for the numerical experiments. In Section \ref{sec:results}, we look into the results of the experiments by measuring and comparing the plasmoid frequency, aspect ratio, and reconnection rate of each simulation case. Finally, in Sect. \ref{sec:discussion} we briefly discuss our results and summarise our conclusions.

\section{Numerical model}
\label{sec:model}

\subsection{Model equations}

The simulations of this paper were performed with the 3D MHD code Bifrost \citep{2011A&A...531A.154G}. This code uses a sixth-order operator for the spatial derivatives and a third-order scheme for the time derivatives, allowing us to minimise the numerical diffusion
due to the discretisation of the equations. In particular, we carried out different 2D simulations focusing on magnetic reconnection at coronal heights. We therefore included Joule heating, viscous heating, and Spitzer conductivity, while excluding radiative heating and cooling terms. Regarding the equation of state, we assume a fully singly ionised ideal gas with a mean molecular weight of 0.616. In addition, gravity is neglected as the whole computational domain lies in
the corona.

\subsection{Resistivity models}

To study reconnection, we employed the three anomalous resistivity models described in the \citetalias{2023A&A...675A..97F} paper, which are summarised below.

\subsubsection{{\hdiff} model}

The {\hdiff} model \citep{2011A&A...531A.154G,1995NordlundGalsgaard} is the default resistivity model of Bifrost. This hyper-diffusive model dynamically scales up the resistivity around gradients in the magnetic field ${\bf B}$ and velocity ${\bf u}$ and can be written as a diagonal tensor, $\bar{\bar{\eta}}_{\mathrm{G11}}$, given by

\begin{align}
    \eta_{\mathrm{G11},xx} &= \frac{\eta_3}{2} \left[
    U_{\mathrm{m},y} \Delta y \mathbb{Q}_y\left( \frac{\partial B_{z}}{\partial y} \right) 
    + U_{\mathrm{m},z} \Delta z \mathbb{Q}_z\left( \frac{\partial B_{y}}{\partial z} \right)
    \right] , \nonumber \\
    \eta_{\mathrm{G11},yy} &= \frac{\eta_3}{2} \left[
    U_{\mathrm{m},z} \Delta z \mathbb{Q}_z\left( \frac{\partial B_{x}}{\partial z} \right) 
    + U_{\mathrm{m},x} \Delta x \mathbb{Q}_x\left( \frac{\partial B_{z}}{\partial x} \right)
    \right] , \nonumber \\    
    \eta_{\mathrm{G11},zz} &= \frac{\eta_3}{2} \left[
    U_{\mathrm{m},x} \Delta x \mathbb{Q}_x\left( \frac{\partial B_{y}}{\partial x} \right) 
    + U_{\mathrm{m},y} \Delta y \mathbb{Q}_y\left( \frac{\partial B_{x}}{\partial y} \right)
    \right] , \nonumber \\ 
    \eta_{\mathrm{G11},xy} &= \eta_{\mathrm{G11},yx} = \eta_{\mathrm{G11},yz} = \eta_{\mathrm{G11},zy}= \eta_{\mathrm{G11},xz} = \eta_{\mathrm{G11},zx} = 0 , \label{eq:eta-hdiff}
\end{align}
\noindent
where
\begin{align}
    U_{\mathrm{m},i} &\equiv  \nu_1 c_\mathrm{f} + \nu_2 |u_i| + \eta_3 \Delta x_i |\nabla_{\perp} u_i| ,\label{eq:diffusive_velocity_Um}\\
    \mathbb{Q}_i(g) &\equiv \frac{\left|  \frac{\partial^2 g}{\partial x_i^2}  \right| \Delta x_i^2}
    {|g| + \frac{1}{q_{\mathrm{max}}} \left|  \frac{\partial^2 g}{\partial x_i^2}  \right| \Delta x_i^2 } ,\label{eq:quenching-operator}
\end{align}
and $c_\mathrm{f} \equiv \sqrt{c_\mathrm{s}^2 + v_\mathrm{A}^2}$, with $c_\mathrm{s}$ and $v_\mathrm{A}$ denoting the sound speed and Alfv\'en speed, respectively. $\nu_1$, $\nu_2$, and $\eta_3$ are free scaling parameters. For this paper, we varied the input value of $\eta_3$ while using fixed $\nu_1=0.03$ and $\nu_2=0.2$, which should be kept as low as possible as discussed in Sect. 3.1.5 of \citetalias{2023A&A...675A..97F}.

\subsubsection{{\syntelis} model}

The {\syntelis} model \citep{2019ApJ...872...32S} applies a scalar resistivity $\eta_{_\mathrm{S19}}$ proportional to the current density ${\bf J}$ as follows: 
\begin{align}
    \eta_{_\mathrm{S19}} &= \left\{ 
    \begin{array}{ll}
        \eta_0, &  |{\bf J}| < J_{\mathrm{crit}} \\
        \eta_0 + \eta_1 |{\bf J}|/J_{\mathrm{crit}}, &  |{\bf J}| \geq J_{\mathrm{crit}} \\
    \end{array}
    \right.  , \label{eq:anomalous_resistivity_syntelis} 
\end{align}
\noindent
where $\eta_0$, $\eta_1$, and $J_{\mathrm{crit}}$ are free parameters. We used $\eta_0 = 3.78\times10^{-2}\ \mathrm{km^2\ s^{-1}}$ and $J_{\mathrm{crit}}=5.00\times10^{-4}\ \mathrm{G\ km^{-1}}$ while varying the input value of $\eta_1$.

\subsubsection{{\yokshi} model}

In the {\yokshi} model \citep{1994ApJ...436L.197Y}, the resistivity $\eta_{_{\mathrm{YS94}}}$ scales with the electron drift velocity $v_\mathrm{d} = |{\bf J}|/(n_\mathrm{e} q_\mathrm{e})$, given the electron density $n_e$ and elementary charge $q_\mathrm{e}$, as follows,
\begin{align}
    \eta_{_{\mathrm{YS94}}} &= \left\{ 
    \begin{array}{ll}
        0, &  v_\mathrm{d} \leq v_\mathrm{c} \\
        \min(\alpha (\frac{v_\mathrm{d}}{v_\mathrm{c}} - 1)^2,\eta_{\mathrm{max}}), &  v_\mathrm{d} > v_\mathrm{c}\\
    \end{array}
    \right. , \label{eq:anomalous_resistivity_yokoyama}
\end{align}
where $v_\mathrm{c}$, $\alpha$, and $\eta_{\mathrm{max}}$ are free parameters. We used $v_{\mathrm{c}} = 8.3\times 10^{-6}\ \mathrm{km\ s^{-1}}$ and $\eta_\mathrm{max}=2000\ \mathrm{km^2\ s^{-1}}$ while varying the input value of $\alpha$.

\subsubsection{Uniform resistivity}

In addition to the three aforementioned anomalous resistivity models, we also used uniform resistivity for comparison purposes,
\begin{align}
    \eta_{_\mathrm{U}} = \eta_0 ,
\end{align}
with various input values for $\eta_0$.

\subsection{Viscosity in Bifrost}

While the resistivity $\bar{\bar{\eta}}$  in our simulations is given by one of the four resistivity models mentioned above, the viscosity tensor $\bar{\bar{\tau}}$ is always given by Bifrost's in-built description, namely
\begin{align}
    \tau_{ij} = \left\{ 
        \begin{array}{ll}
            \rho \Delta x_i U_{{\rm v}, i} \frac{\partial u_i}{\partial x_i} \mathbb{Q}_i\left(\frac{\partial u_i}{\partial x_i}\right), & i=j \\
            \rho \left[\Delta x_j U_{{\rm v}, j} \frac{\partial u_i}{\partial x_j} \mathbb{Q}_j\left(\frac{\partial u_i}{\partial x_j}\right) +  \Delta x_i U_{{\rm v}, i} \frac{\partial u_j}{\partial x_i} \mathbb{Q}_i\left(\frac{\partial u_j}{\partial x_i}\right)\right], & i \neq j,
        \end{array}
    \right.
\end{align}
where
\begin{align}
    U_{\mathrm{v},i} &\equiv  \nu_1 c_\mathrm{f} + \nu_2 |u_i| + \nu_3 \Delta x_i |\nabla u_i| ,\label{eq:diffusive_velocity_Uv}
\end{align}
and $\nu_3$ is a free scaling parameter, which is set to 0.3 in our simulations.

\subsection{Model setup}

\begin{figure}
    \centering
    \includegraphics[width=\columnwidth]{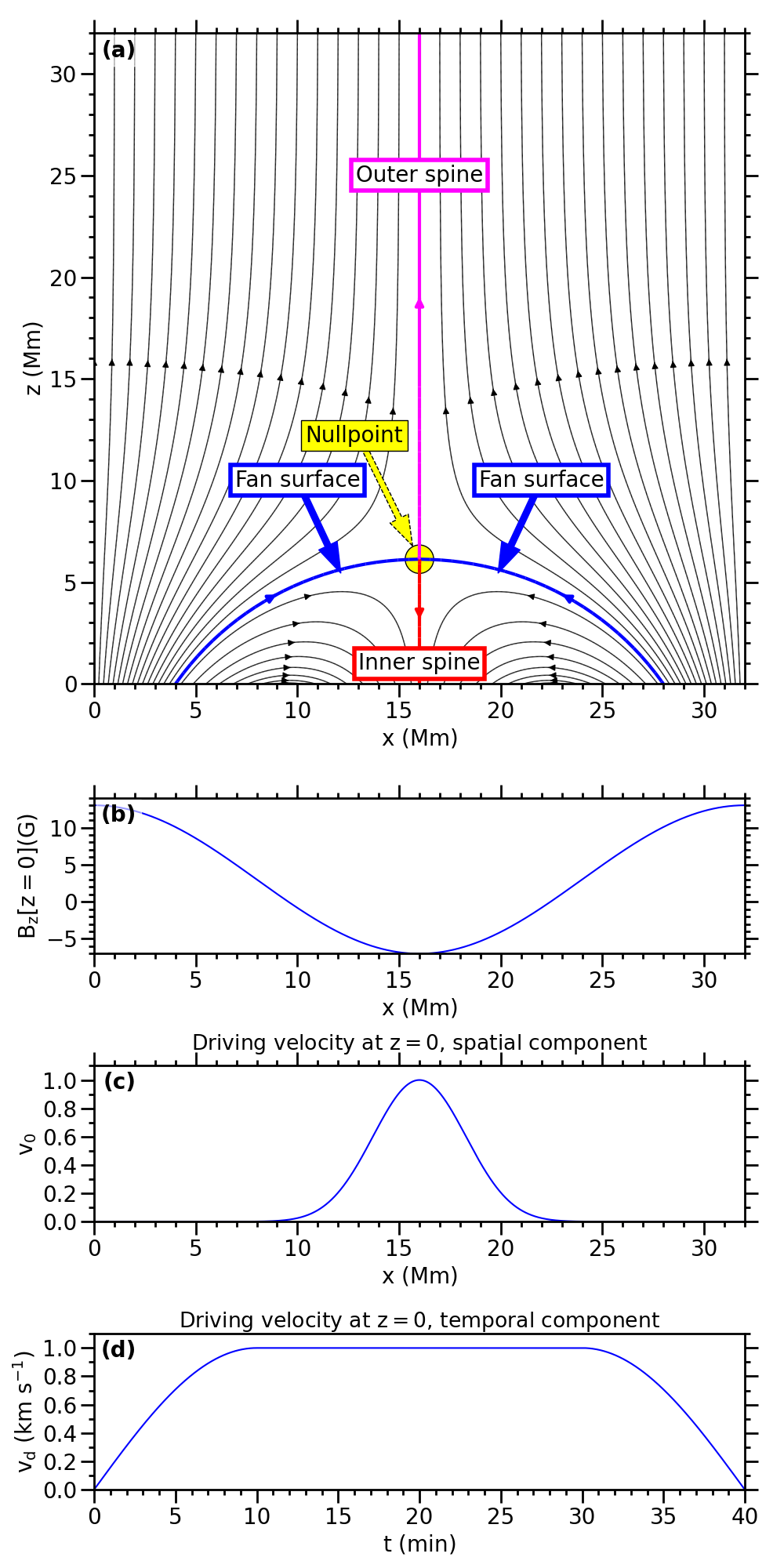}
    \caption{Model setup. Panel (a) shows the initial magnetic field topology. Panel (b) displays the vertical component of this field measured at $z=0$. The inner spine of the magnetic field topology is moved in positive $x$-direction with a driving velocity given by a product of a spatial factor, plotted in panel (c), and a temporal factor, plotted in panel (d).}    
    \label{fig:model-setup}
\end{figure}

\begin{table*}[t!]
    \caption{Simulations with their respective grid points, resistivity model, and resistivity peak values, $\eta_\mathrm{p}\ (\mathrm{km^2\ s^{-1}})$.}
    \label{tab:resistivity-cases}
    \centering
    \setlength\doublerulesep{0.5pt}
    \begin{tabular}{cccl}
        \hhline{====}
        \rule{0pt}{2.0ex}
        Group & Resolution & Resistivity model & $\eta_\mathrm{p}\ (\mathrm{km^2\ s^{-1}})$ \\
        \hline
        \rule{0pt}{2.5ex}
        S1-5 & $2048\times 2048$ & {\syntelis} & 
        125, 87.3, 62.4, 42.7, 24.8
        \\
        Y1-8 & $2048\times 2048$ & {\yokshi} & 
        
        155, 89.2., 71.7, 57.4, 43.9, 34.2, 26.7, 21.7
        \\
        G1-7 & $2048\times 2048$ & {\hdiff} & 
        650, 309, 211, 131, 68.3, 36.1, 19.9
        \\
        U1-4 & $2048\times 2048$ & Uniform ($\eta = \eta_\mathrm{p}$) & 75.6, 37.8, 18.9, 15.1  \\
        4kU1-9 & $4096\times 4096$ & Uniform ($\eta = \eta_\mathrm{p}$) & 75.6, 37.8, 18.9, 15.1, 7.56, 3.78, 1.89, 0.945, 0.473  \\
        8kU1-11 & $8192\times 8192$ & Uniform ($\eta = \eta_\mathrm{p}$) & 75.6, 37.8, 18.9, 11.3, 7.56, 3.78, 
        1.89, 0.945, 0.473, 0.378, 0.189 \\
        \hline
    \end{tabular}
    
\end{table*}

For the initial condition, we imposed a 2D fan-spine topology in a similar fashion to \cite{2019A&A...628A...8P} and \cite{2022ApJ...935L..21N}. In particular, the horizontal and vertical components of the magnetic field are respectively given by
\begin{align}
    B_x &= B_1 e^{-kz} \sin (kx) \label{eq:init-bx}, \\
    B_z &= B_0 + B_1 e^{-kz}  \cos (kx) \label{eq:init-bz}, 
\end{align}
where $B_1 = 10$ G, $k = \pi/16\ \mathrm{Mm^{-1}}$, and $B_0 = 3$ G. The external field $B_0$ was set to resemble that of a typical quiet-Sun coronal hole \citep{2019A&A...629A..22H}. Panels (a) and (b) of  Fig.~\ref{fig:model-setup} contain the initial magnetic field topology and $B_z(x,z=0)$, respectively. These panels show that the imposed field has a negative parasitic polarity in a positive background, which leads to a null-point at $z=6.13$ Mm.
The initial temperature and mass density were uniformly set to $T_0 = 0.61$ MK and $\rho_0 = 3\times 10^{-16}\ \mathrm{g\ cm^{-3}}$ to resemble typical values of the lower corona. 

Concerning the boundary conditions, the side boundaries were periodic. The top boundary was treated by an absorbing layer on all MHD variables in order to ensure that any wave that hits the boundary is not reflected. At the bottom boundary, an absorbing layer was applied on the mass density $\rho$, internal energy density $e$, and the vertical velocity $u_z$. For the horizontal velocity $u_x$, a driving condition was imposed to move the inner spine of our fan-spine topology with a velocity up to 1 km s$^{-1}$. More specifically, $u_x$ is a product of two components, similar to \citet{2019A&A...628A...8P}, defined as
\begin{align}
    u_x(x,z=0,t) = v_\mathrm{d}(t) v_0(x) \label{eq:bound-ux}.
\end{align}
The spatial component $v_0(x)$ is given by
\begin{align}
    v_0(x) = \left( \frac{1+\cos\left(\pi(x-L_x)/L_x\right))}{2} \right)^{10} , \label{eq:bound-ux-spatial}
\end{align}
where $L_x = 16$ Mm, which is the half-width of the computational domain. The temporal component $v_d(t)$ is as follows
\begin{align}
    v_\mathrm{d}(t) = v_\mathrm{p} \left\{ 
    \begin{array}{ll}
         \sin\left(0.5\pi t/t_\mathrm{r}\right) & t \in [0, t_\mathrm{r}] \\
         1.0 & t \in [t_\mathrm{r}, t_\mathrm{d}-t_\mathrm{r}] \\
        \sin\left(0.5\pi(t_\mathrm{d}-t)/t_\mathrm{r}\right) & t \in [t_\mathrm{d} - t_\mathrm{r}, t_\mathrm{d}] 
    \end{array}
    \right. , \label{eq:bound-ux-time}
\end{align}
with a peak velocity of $v_\mathrm{p} = 1\ \mathrm{km\ s^{-1}}$, a ramping time of $t_\mathrm{r} = 10$ min, and a total driving time of $t_\mathrm{d} = 40$ min. The spatial and temporal components of this driving velocity are shown in panels (c) and (d) of Fig.~\ref{fig:model-setup}. The magnetic field at the bottom boundary is line-tied to the flow. This was ensured by setting the magnetic field in the ghost zones to be anti-symmetric around the boundary value. The same anti-symmetric-around-boundary-value condition was applied on $u_x$ in the ghost zones.

The numerical experiments span a $32 \times 32\ \mathrm{Mm^2}$ physical domain and were run for 40 min. In particular, we performed 44 different simulations grouped as follows: (1) the 2k simulations, that is, 24 cases with a resolution of $2048 \times 2048$ grid points, using either uniform, {\syntelis}, {\yokshi}, or {\hdiff} resistivity with various input values for the scaling parameters; (2) the 4k simulations, that is, nine experiments, each with a resolution of $4096 \times 4096$ grid points, all using a uniform resistivity with different values of $\eta_0$; and (3) the 8k simulations, that is, 11 runs with an $8192 \times 8192$ resolution, also using a uniform resistivity with different values of $\eta_0$.
The details of all the cases are listed in Table~\ref{tab:resistivity-cases}; models are labelled with a letter, which denotes the chosen resistivity model, and a number that decreases with increasing resistivity.
The fourth column displays the peak value $\eta_{\rm p}$, the meaning of which is as follows. For the uniform-resistivity cases, $\eta_{\rm p}$ is equal to the uniform value $\eta_0$. For any of the anomalous resistivity cases (S1-5, Y1-8, G1-7), $\eta_{\rm p}$ denotes the maximum resistivity in the current sheet averaged over the time period $t\in[15,35]$ min and is directly proportional to the input value of the scaling parameter of the resistivity model applied in the given case.

In the 2k simulations, the scaling parameter for each resistivity model varied from the minimum required for stability up to 1-2 orders of magnitude above, or to a level that entirely prevents plasmoid formation (resulting in a few cases of steady reconnection). Similar variations were applied in the 4k and 8k simulations. Notably, in these cases, the resistivity could be set considerably lower than in the 2k simulations while maintaining stability. On the other hand, if the resistivity terms are completely removed,  the simulations become numerically unstable. This fact indicates that the numerical diffusion  due to the discretisation of the equations is negligible with respect to the explicit resistivity terms in the small regions with large gradients or jumps in the variables, as in current sheets.


\section{Results}
\label{sec:results}

\begin{figure}
    \centering

    \includegraphics[width=\columnwidth]{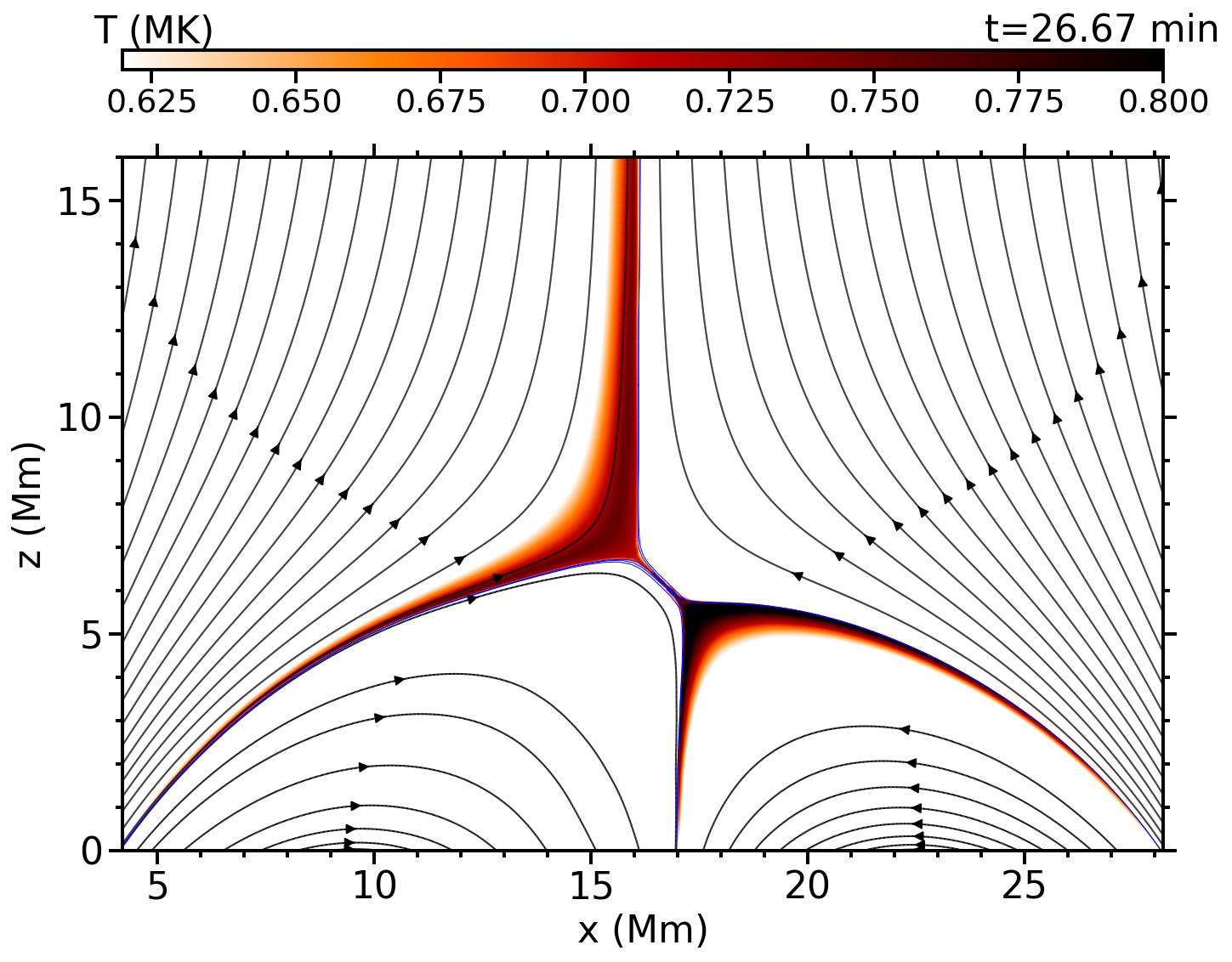}

    \caption{Temperature and magnetic field topology taken from simulation case 8kU6. A movie of the time evolution of the map for $t\in[0,40]$ min is available online.}
    \label{fig:Results_tg_40min}
\end{figure}

\begin{figure*}
    \centering

    \includegraphics[width=\textwidth]{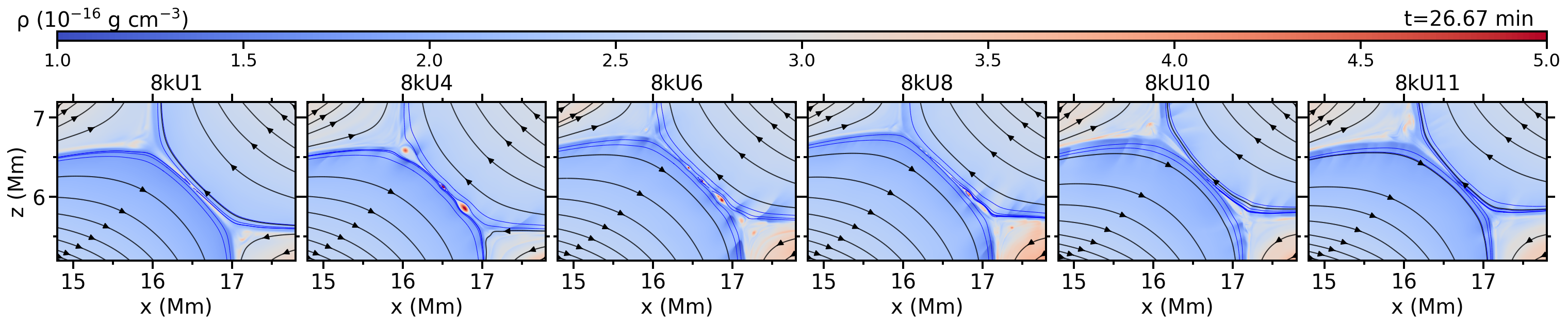}

    \caption{Mass density and magnetic field topology around the current sheet for six of the 8k simulation cases. A movie of the time evolution of the maps for $t\in[25,30]$ min is available online to show how plasmoids and shocks originate along the current sheet.}
    \label{fig:Results-rho-zoom}
\end{figure*}

\subsection{Overview}

In all simulations, the inner spine undergoes a positive $x$-directional displacement due to the boundary driving velocity. As a consequence, the null-point collapses, leading to a tilted current sheet between the inner and outer spine. 
Following the behaviour of the driver, the length of the current sheet increases during the first 15 minutes of the simulation; it then remains roughly constant for 20 minutes before finally decreasing during the final 5 minutes of the simulations.
At the current sheet, reconnection occurs continuously, significantly heating the plasma. 
As a representative example, Fig.~\ref{fig:Results_tg_40min} shows the temperature of the 8kU6 case at $t=26.7$~min with the magnetic field topology superimposed.
An animation of the full time evolution of the map is available online.
In all the simulations, the temperature profile has roughly the same shape as shown in the image, albeit with distinct peak temperatures, which range from 0.72 to 0.83 MK.

The differences between the simulations are more evident regarding other physical quantities such as mass density, which is displayed in Fig.~\ref{fig:Results-rho-zoom} at $t=26.7$ min for six of the 8k cases (see also associated animation). For instance, in case 8kU1,
no evident plasmoids are seen, while plasmoids appear
frequently in the other cases, moving in either direction along the current sheet. In some cases, several plasmoids merge together, a phenomenon referred to as coalescence instability \citep{1977PhFl...20...72F}. In the following, we analyse the characteristics of the reconnection in all simulation cases listed in Table~\ref{tab:resistivity-cases}.

\begin{figure}
    \centering
    \includegraphics[width=\columnwidth]{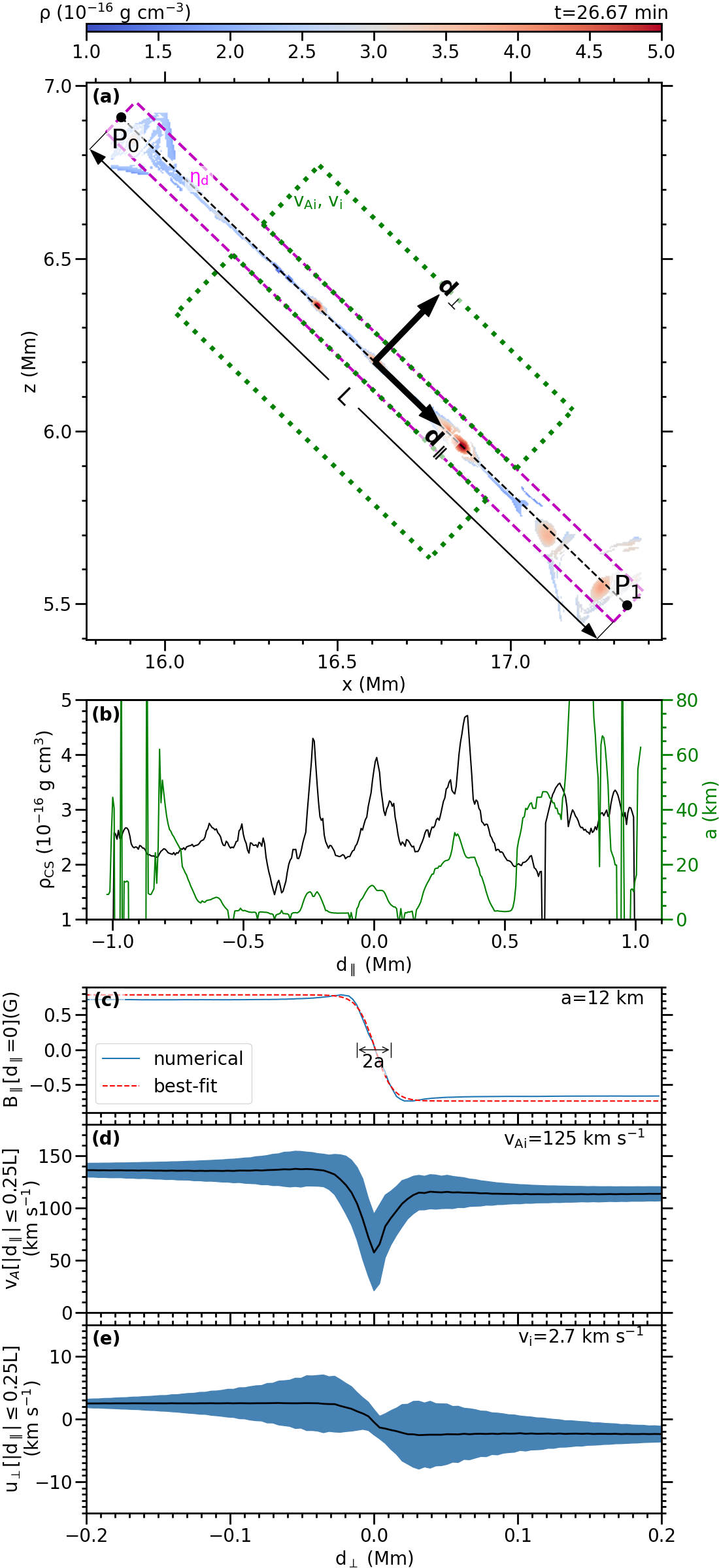}
    \caption{Characteristics of the current sheet (case 8kU6). Panel (a): Mass density $\rho$ in the current sheet, mapped for $L_B\leq 20$ km. The dashed line with endpoints $P_0$ to $P_1$ marks the current sheet, with coordinate axes for $d_\parallel$ and $d_\perp$ plotted in. Diffusion and inflow regions are delimited by magenta and green rectangles, respectively. Panel (b): Average density $\rho_{CS}$ (black curve) and width $a$ (green curve) of the current sheet. Panel (c): Parallel component of magnetic field, $B_\parallel$ (blue), across the current sheet and the best-fit (red) curve used to estimate $a$ at $d_\parallel=0$. Panels (d) and (e): Alfven velocity $v_A$ (d) and perpendicular velocity $u_\perp$ (e) across the current sheet. Blue area maps the ranges of all values for $|d_\parallel|\leq 0.25L$, and black curve plots the average. Estimated inflow region mean values are printed in top right corner.}
    \label{fig:measuring-cs-width}
\end{figure}

\begin{figure}
    \centering
    \includegraphics[width=\columnwidth]{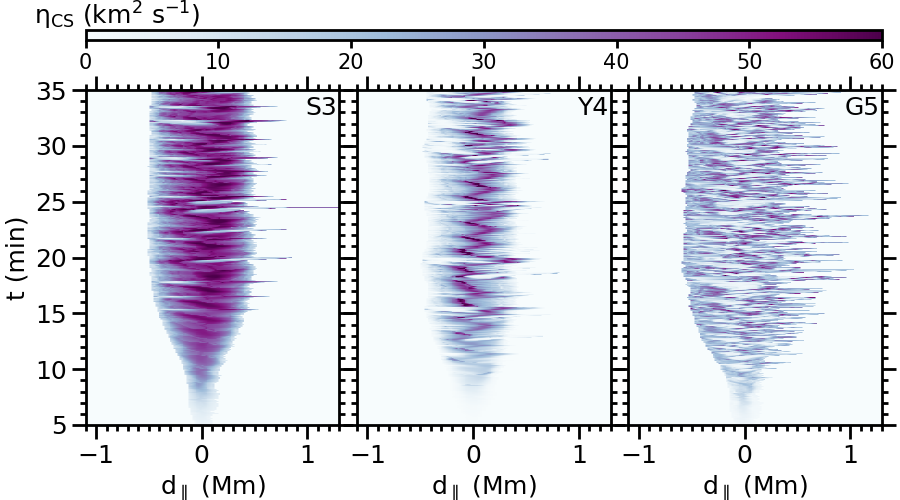}
    \caption{Evolution of the resistivity $\eta_\mathrm{CS}$ along the current sheet for the S3, Y4, and G5 models. The three cases shown are 2k simulations with anomalous resistivity with $\eta_\mathrm{p} = 60$ km$^2$ s$^{-1}$.}
    \label{fig:eta_plots}
\end{figure}

\subsection{Data analysis method}

\subsubsection{The current sheet}

As a first step of our analysis, we define our current sheet as the region with a characteristic length $L_B \equiv (|{\bf J}|/|{\bf B}|)^{-1} \leq 20$ km, filtering away any cells that belong to the spines and fan surfaces. The  $L_B \leq 20$ km threshold ensures that we consider elements with a characteristic length larger than the lowest resolution we have ($\Delta x = \Delta z = 15.6$ km in the 2k cases). As an example, the top panel of Fig.~\ref{fig:measuring-cs-width} contains a density map within the current sheet in the 8kU6 simulation. The current-sheet axis is found through a linear fit of the cells fulfilling the aforementioned condition, and the current-sheet length, $L$, is then measured as the distance between its extremes, labelled $P_0$ and $P_1$, as shown in the figure. Having located the current-sheet axis, we define a coordinate system centred at the middle of the current sheet, using the distances along ($d_\parallel$) and perpendicular to the current sheet ($d_\perp$); see Fig.~\ref{fig:measuring-cs-width} for coordinate axes. 

To measure the current-sheet width, we projected the magnetic field onto the coordinate system of the current sheet. Its component parallel to the sheet, $B_\parallel$, has a \citet{1962NCim...23..115H} current sheet-like profile in its variation with $d_\perp$, having nearly oppositely equal values on each side of the sheet.
We therefore found the current-sheet width $a$ by fitting  $B_{\parallel}$ with a hyperbolic tangent.
Panel (c) of Fig.~\ref{fig:measuring-cs-width} depicts the method, showing  $B_{\parallel}$ (blue curve) at $d_\parallel=0$, and its fit $B_\mathrm{fit}$ (red curve) as functions of $d_{\perp}$. The variation of the width  along the current sheet is given in Panel (b) (green curve). The large peaks in this curve correspond to plasmoids, as evidenced by the density variations along the current sheet ($\rho_\mathrm{CS}$) shown in black in the same panel.
In subsequent sections, we use the average width over the whole current sheet  $\bar{a}$ to estimate the inverse aspect ratio $\bar{a}/L$ (Sect.~\ref{sec:aspect}), as well as density variations along the current sheet to measure the frequency of plasmoids (Sect.~\ref{sec:plasmoids}).

To illustrate how the different anomalous resistivity models work on the current sheet, Fig.~\ref{fig:eta_plots} maps the resistivity $\eta_\mathrm{CS}$ along the current sheet for three 2k simulation cases (S3, Y4, and G5), which all reach a peak value of around $60\ \mathrm{km^2 s^{-1}}$.
The resistivity of S3 has a weaker variation along the current sheet than the other two cases here, which is due to the fact that the resistivity of the {\syntelis} model is only linearly proportional to current density. Therefore, one might expect the results of this resistivity model to lie closer to those of uniform resistivity (for the same resolution). Case G5, on the other hand, shows by far the most variation in the resistivity along the sheet out of these three cases; this is due to the more dynamic behaviour of the {\hdiff} model.

\subsubsection{The diffusion region}
The diffusion region of the reconnection site was defined as the region around the current sheet delimited by $|d_\parallel| \leq 0.50L$ and $|d_\perp| \leq 60\ \mathrm{km}$, marked by a magenta dashed rectangle in the top panel of Fig~\ref{fig:measuring-cs-width}. We chose to set the diffusion region half-width to 60 km for two reasons: (a) this threshold is slightly bigger than the peak value of the sheet width $a$  measured in the largest plasmoids in our simulation cases, and (b) it ensures  that the magnetic Reynolds number $Re \equiv L_B |{\bf u}|  /\eta$ is always larger than 100 outside this region. Thus, this diffusion region marks the area where the resistivity has a significant effect on the plasma. The mean resistivity of the diffusion region, $\eta_\mathrm{d}$, is used when estimating the effective Lundquist number.

\subsubsection{The inflow regions}
The inflow regions of the reconnection site were defined as the areas delimited by $|d_\parallel| \leq 0.25L$ and $60\ \mathrm{km} \leq |d_\perp| \leq 300\ \mathrm{km}$, marked by green dotted rectangles in the top panel of Fig~\ref{fig:measuring-cs-width}. This threshold ensures that the inflow regions lie just outside the diffusion region (so $Re>100$), and the Alfvén speed here is more or less constant with distance from the sheet. The delimitation of $|d_\parallel| \leq 0.25L$ is to avoid the areas near the endpoints of the current sheet where the Alfvén speed fluctuates more rapidly. 

With this definition, the inflow Alfvén speed $v_{Ai}$ was measured as the mean Alfvén speed within the green dotted rectangles. Similarly, the inflow velocity $v_i$ was measured as the mean absolute value of the velocity $u_\perp$ perpendicular to the current sheet within the inflow region. 
In panels (d) and (e) of Fig.~\ref{fig:measuring-cs-width}, we show both quantities as a function of $d_\perp$. The black curve plots the average values taken over $|d_\parallel|\leq 0.25L$, while the blue area shows the ranges within one standard deviation. The estimated (equilibrium) values for the inflow Alfvén speed and the inflow velocity (at a given time and for a given case) is computed as the mean value of these black curves for $0.06\ \mathrm{Mm} \leq |d_\perp| \leq 0.3\ \mathrm{Mm}$, which is printed in the upper right corners of the panels. 

Finally, the reconnection rate $M_\mathrm{Ai}$ in each simulation case can be estimated as the mean of $v_\mathrm{i}/v_\mathrm{Ai}$, which is analysed in Sect.~\ref{sec:reconection-rate}. Similarly, the effective Lundquist number $S_\mathrm{L}$ is estimated as the mean of $L v_\mathrm{Ai} / \eta_\mathrm{d}$, which is a central part of the analysis in the following sections. For both quantities, the mean values are time averages over $t \in [15, 35]$ min due to the fact that the current-sheet length is approximately stable during that time period.

\subsection{Frequency of plasmoids along the current sheet}
\label{sec:plasmoids}

\begin{figure*}[h!]
    \centering
    \includegraphics[width=\textwidth]{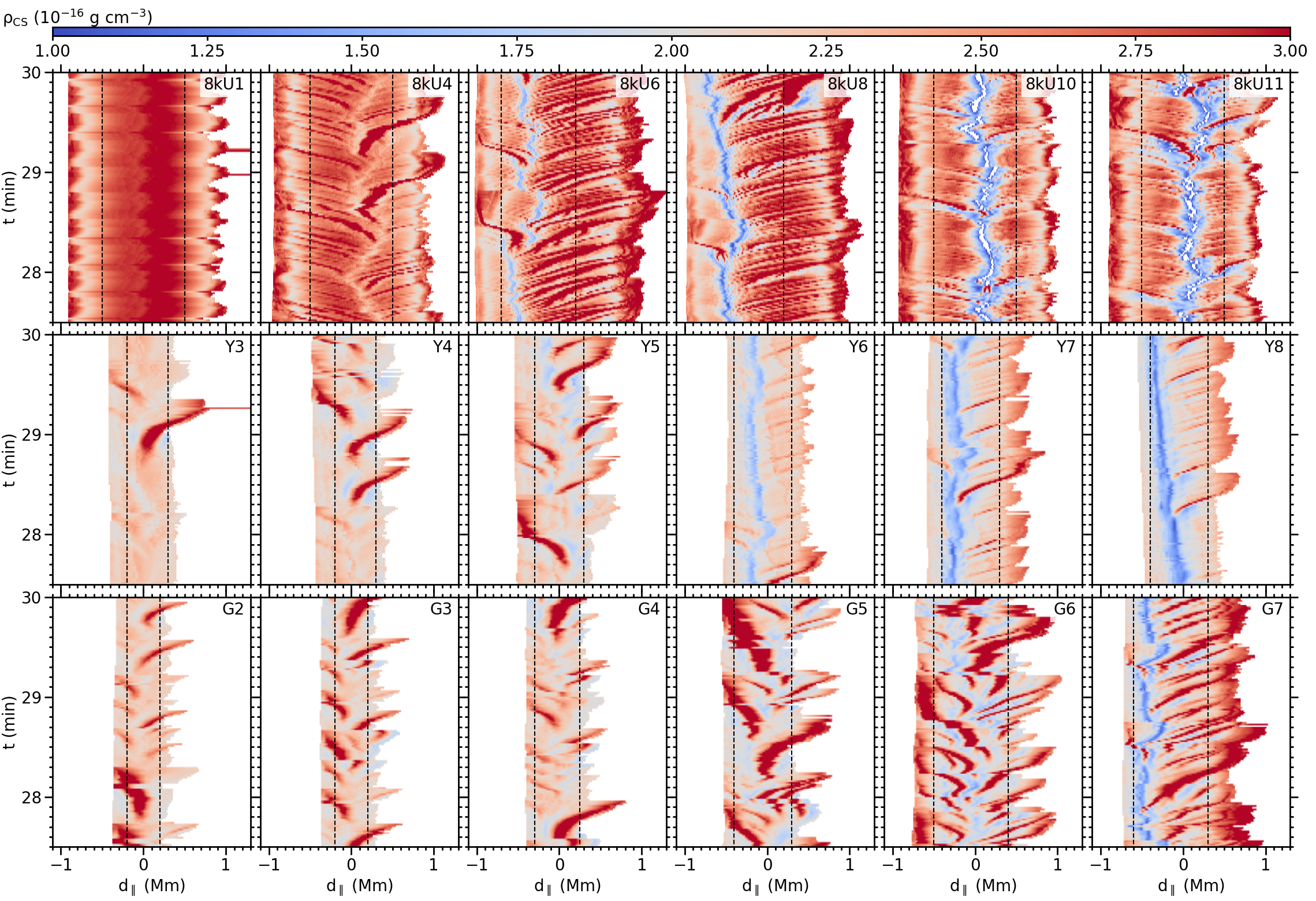}
    \caption{Evolution of mass density $\rho_\mathrm{CS}$ along the current sheet over time for selected simulation cases. Dashed lines mark the locations where a peak detection algorithm was used to count the number of plasmoids occurring per time.}
    \label{fig:rho_against_length_time}
\end{figure*}

\begin{figure*}[h!]
    \centering
    \includegraphics[width=\textwidth]{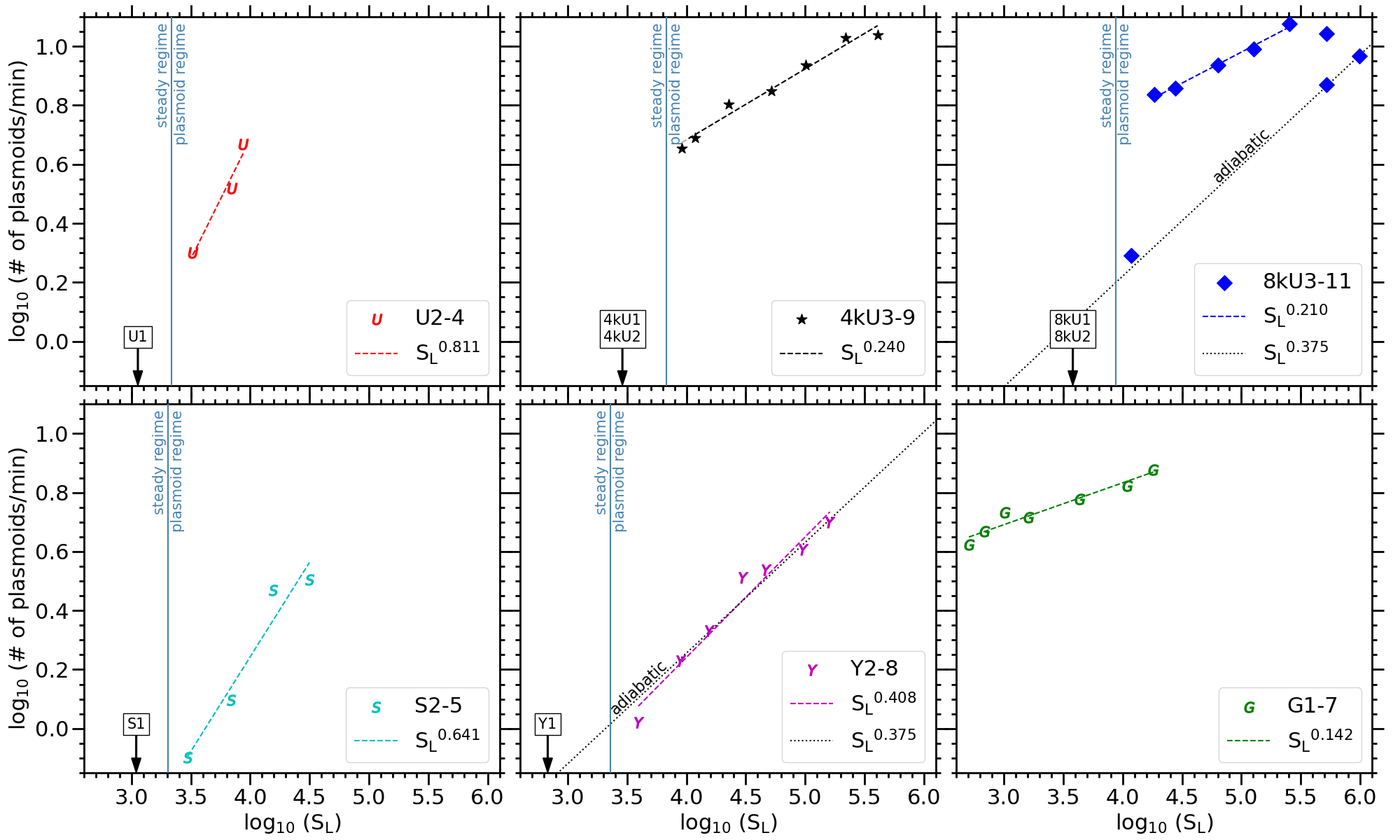}
    \caption{Plasmoid frequency, measured as the number of plasmoids generated along the current sheet per minute, plotted against Lundquist number $S_\mathrm{L}$ for each simulation case. Results are displayed for cases with uniform resistivity (top panels, with 2k, 4k, and 8k cases in separate panels) and anomalous resistivity (bottom panels, with {\syntelis}, {\yokshi}, and {\hdiff} cases in separate panels). For cases within a certain range of Lundquist numbers, the plasmoid frequency scales roughly with Lundquist number by a power law $S_\mathrm{L}^p$, and best-fit curves for these cases are plotted as dashed lines in each panel. The curve for the adiabatic power law $S_\mathrm{L}^{0.375}$ is plotted as a dotted line for the cases where a nearly adiabatic scaling between plasmoid number and Lundquist number occur. A vertical line marks the Lundquist number below which steady reconnection occurs and above which plasmoid-mediated reconnection occurs.
    Cases U1, 4kU1-2, 8kU1-2, S1, and Y1 have no plamsoids, as indicated by the label placed inside the steady-reconnection regime.}
    \label{fig:plasmoid_lundquist_relation}
\end{figure*}

The frequency of plasmoids in the current sheet is studied  here through the variation in mass density $\rho_\mathrm{CS}$ measured along the sheet (Fig.~\ref{fig:measuring-cs-width}, panel b), which, for our case, was found to be easier than detecting null-points following the method described by \cite{2010PhPl...17f2104H}. To demonstrate this,  $\rho_\mathrm{CS}$ is mapped against $d_\parallel$ and time in Fig.~\ref{fig:rho_against_length_time} for (top panels) the same six 8k simulation cases as in Fig.~\ref{fig:Results-rho-zoom}, along with six 2k cases with the {\yokshi} resistivity (middle panels) and six with the {\hdiff} resistivity (bottom panels). Plasmoids are here identified as dark red stripes tilted either upwards to the left or to the right, depending on which way the plasmoids move along the current sheet. In agreement with the movie of Fig.~\ref{fig:Results-rho-zoom}, no plasmoids appear in case 8kU1. On the other hand, plasmoids appear
frequently in cases 8kU4, 8kU6, and 8kU8. In case 8kU4, a roughly equal number of plasmoids move upwards to the left along the current sheet as those moving downward to the right, while in cases 8kU6 and 8kU8, the majority move in the latter direction. In cases 8kU10 and 8kU11, the dark stripes are very thin and barely visible, which indicates that most of the plasmoids have diminished and in such a way that they are only visible as outward-propagating shocks, which is also seen in the movie.
These cases are not perfectly shock-mediated, as plasmoids still occur (though the larger plasmoids occur only rarely here), but they are significantly closer to the shock-mediated regime than cases 8kU4-9.
Therefore, the 8k cases seemingly cover three different types of reconnection: steady (Sweet-Parker-like), plasmoid-mediated, and (nearly) shock-mediated (Petschek-like) reconnection. In all 8k cases, the current-sheet length, as measured in the figure as the width of the coloured region, lies roughly around 2 Mm. The corresponding maps for the 4k cases (not shown in the figure) appear very similar to the 8k cases, though with a slightly shorter current-sheet length. Similar plasmoid patterns are also found in the 2k uniform resistivity cases for a narrower range of Lundquist numbers.

Among the {\yokshi} resistivity cases (see Fig.~\ref{fig:rho_against_length_time}, middle row), the number of plasmoids (as seen as the dark stripes in the maps) clearly increases from Y3 to Y5. The plasmoids are more difficult to detect by eye in cases Y6-8, but a closer look reveals a significant number of very thin stripes. Hence, the plasmoids as reproduced with the {\yokshi} model seem to diminish in size (but not necessarily in number) as the resistivity gets sufficiently low. This indicates that reconnection reproduced with this resistivity model may approach steady Petschek reconnection ---which is characterised by shocks instead of plasmoids--- as the resistivity decreases. In all of the {\hdiff} cases, the plasmoids are relatively large in size, and are clearly more numerous in the lower-resistivity cases (especially in G5-7) than in the higher-resistivity cases. Among the {\syntelis} cases, which are not shown in the figure, a minor decrease in plasmoid size is seen from cases S4 to S5, similar to that of the {\yokshi} cases, but of a lesser degree. All the 2k cases have a shorter current sheet than the 8k (and 4k) cases, which is due to a higher numerical diffusion that sets a stricter limit on the current-sheet length. The current-sheet length in the {\hdiff} cases increases as the resistivity decreases, in agreement with the discussion in Sect. 3.1.5 of \citetalias{2023A&A...675A..97F} on how current-sheet length depends on the scaling of the anomalous resistivity models. A similar but weaker scaling between current-sheet length and resistivity is found in the {\yokshi} and {\syntelis} cases.

In order to measure the frequency of plasmoids for each simulation case, we picked specific locations along the current sheet where we measure the density as a function of time. These locations are marked with dashed vertical lines in each panel of Fig.~\ref{fig:rho_against_length_time}. For most of the cases, plasmoids move in either direction, and so we picked two locations for measuring the density curves. These locations were picked in such a way that each plasmoid passes through one of the locations, but not both. Plasmoids passing through one of those points are then detected as spikes in the density curves. Hence, the total number of plasmoids generated along the current sheet is given by the total number of spikes in the density curves. In the shock-mediated cases 8kU10-11 and Y6-8, the shocks are also seen as spikes in these curves.

The frequency of plasmoids for the different simulation cases ---measured as the total number of plasmoids found in each case in the time interval $t\in [15, 35]$ min divided by 20 min--- is plotted against Lundquist number in Fig.~\ref{fig:plasmoid_lundquist_relation}. The results are grouped into different panels by resistivity model and resolution. For a certain range of Lundquist number within each group of cases, the plasmoid frequency increases roughly with Lundquist number by a power law $S_\mathrm{L}^p$, and we used curve fitting to find the best-fitting value of $p$, and the best-fit curves are plotted as dashed lines. For the shock-mediated cases, we use the term `shock frequency' instead of `plasmoid frequency', as the majority of the spikes found in the density curves in those cases are seen only as shocks propagating out of the reconnection site.

Among the uniform resistivity cases, as seen in the top panels of Fig.~\ref{fig:plasmoid_lundquist_relation}, cases U1, 4kU1-2, and 8kU1-2 follow steady reconnection, and therefore no plasmoids occur, as indicated by their label placed to the left of the vertical blue line in each panel. The other cases are plasmoid-mediated or shock-mediated (8kU10 and 8kU11). As for the 2k cases, plasmoid-mediated reconnection is reproduced only for a narrow range of Lundquist numbers given by $3.5 \leq \log S_\mathrm{L} \leq 4.0$, below which steady reconnection occurs, and above which numerical instability occurs. Within the plasmoid-mediated regime, given by cases U2-U4, the plasmoid frequency ranges from 2.0 to 4.7 plasmoids per minute, with a scaling with Lundquist number given by $S_\mathrm{L}^{0.811}$, which is much stronger than the  $S_\mathrm{L}^{0.375}$ scaling found by \citet{2007PhPl...14j0703L} for an adiabatic medium. Regarding the plasmoid-mediated 4k cases (4kU3-9), the plasmoid number ranges from 4.5 to 11 plasmoids per minute for Lundquist numbers of $3.9 \leq \log S_\mathrm{L} \leq 5.6$ with a scaling of $S_\mathrm{L}^{0.240}$, which is weaker than the above-mentioned adiabatic scaling, and is relatively close to the $S_\mathrm{L}^{0.223}$ scaling found in the non-adiabatic cases of \citet{2022A&A...666A..28S}. Regarding the 8k cases, the plasmoid frequency ranges from 6.9 to 12 plasmoids per minute for Lundquist numbers of  $4.2 \leq \log S_\mathrm{L} \leq 5.4$ with scaling of $S_\mathrm{L}^{0.210}$, which is even weaker than the scaling of the plasmoid-mediated 4k cases and is even closer to the scaling of \citet{2022A&A...666A..28S}. In the shock-mediated cases 8kU10 and 8kU11, the measured frequency of shocks is lower than the plasmoid frequencies of 8kU7-9. These two cases fit well to the (dotted) line for the  $S_\mathrm{L}^{0.375}$ scaling, indicating that the frequency of shocks generated in this type of (Petschek-like) reconnection scales adiabatically with Lundquist number. Case 8kU3 is seemingly in an intermediate state between the steady-reconnection regime and the plasmoid-mediated regime, and case 8kU9 is in an intermediate state between the plasmoid-mediated and shock-mediated regimes. By comparing the results for uniform resistivity with the three different resolutions, we see that the plasmoid frequency tends to converge towards higher values with a weaker scaling with Lundquist number as the resolution is increased. The difference is smaller between the 4k and 8k cases than between the 2k and 4k cases.

Among the 2k cases with the {\syntelis} resistivity model (bottom left panel), steady reconnection occurs for $\log S_\mathrm{L} < 3.4$ (case S1). For $3.4 \leq \log S_\mathrm{L} \leq 4.5$ (cases S2-5), the plasmoid frequency ranges from 0.8 to 3.2 plasmoids per minute with a scaling of $S_\mathrm{L}^{0.641}$, a significantly stronger scaling than the adiabatic one, though weaker than the 2k cases with uniform resistivity (for higher Lundquist number, numerical instability occurs). Among the {\yokshi} cases (bottom centre panel), Y1 has steady reconnection, and in cases Y2-8, the plasmoid frequency (or shock frequency for Y6-8) ranges from 1.1 to 5.0 plasmoids (or shocks) per minute for $3.6 \leq \log S_\mathrm{L} \leq 5.2$ with a scaling of $S_\mathrm{L}^{0.408}$. With the (dotted) line for adiabatic scaling $S_\mathrm{L}^{0.375}$ added to the panel, we see that the {\yokshi} resistivity model is capable of reproducing a nearly adiabatic scaling between plasmoid (or shock) frequency and Lundquist number. As for the {\hdiff} cases (bottom right panel), the plasmoid frequency ranges from 4.2 to 7.6 plasmoids per minute for $2.6 \leq \log S_\mathrm{L} \leq 4.3$ with a scaling of $S_\mathrm{L}^{0.142}$. Therefore, with the 2k resolution, the {\hdiff} model reproduces the highest plasmoid frequency with the weakest scaling to Lundquist number. Moreover, the {\hdiff} cases are the only 2k cases where plasmoid frequency is found to scale more weakly with Lundquist number than the adiabatic scaling, and is closer to the scaling of \citet{2022A&A...666A..28S} than the other 2k cases.

The key 
 findings of this plasmoid analysis are as follows: we observe that with uniform resistivity and a sufficiently high resolution (4k and 8k cases), the dependency between plasmoid formation and Lundquist number may be divided into three regimes: (1) a steady-reconnection regime, for Lundquist numbers lower than $10^4$; (2) a plasmoid-mediated regime with a subadiabatic scaling between plasmoid number and Lundquist number similar to that of \cite{2022A&A...666A..28S} for Lundquist numbers between  roughly $10^4$ and $4\times 10^5$; and (3) a shock-mediated regime for Lundquist numbers above roughly $4\times 10^5$, where the frequency of shocks follows an adiabatic scaling with Lundquist number similar to that predicted by \cite{2007PhPl...14j0703L}. With uniform resistivity, very high resolution (as in our 8k cases, $\Delta x = \Delta z = 3.9$ km) is needed to obtain numerically stable simulations with a Lundquist number high enough to reproduce the latter, shock-mediated regime.

 For lower resolutions (as in our 2k cases, $\Delta x = \Delta z = 15.6$ km), uniform resistivity is not a suitable resistivity model for studying plasmoid formation, as plasmoid-mediated reconnection is reproduced only within a narrow range of Lundquist numbers (between $3\times 10^3$ and $10^4$) without breaking numerical stability along the current sheet. Within this range, the plasmoid number increases rapidly with Lundquist number. The {\syntelis} resistivity model allows numerically stable simulations with plasmoid-mediated reconnection for a slightly wider range, but still with a significantly strong scaling between plasmoid number and Lundquist number.  The {\yokshi} model is capable of reproducing plasmoid- or shock-mediated reconnection for a relatively wide range of Lundquist numbers and shows an almost perfectly adiabatic scaling between plasmoid or shock frequency and Lundquist number. The {\hdiff} model is capable of reproducing plasmoid frequencies closer to those seen in the high-resolution high-$S_L$ cases (G7 having $\sim 7.6$ plasmoids per minute, and the 8kU4-8 having about 7-12 plasmoids per minute), and the scaling between plasmoid number and Lundquist number is weaker than in the adiabatic case, which is in fair agreement with the scaling seen in our higher-resolution cases as well as with the scaling found by \cite{2022A&A...666A..28S}.

\subsection{Aspect ratio of the current sheet}
\label{sec:aspect}

\begin{figure*}[h!]
    \centering
    \includegraphics[width=\textwidth]{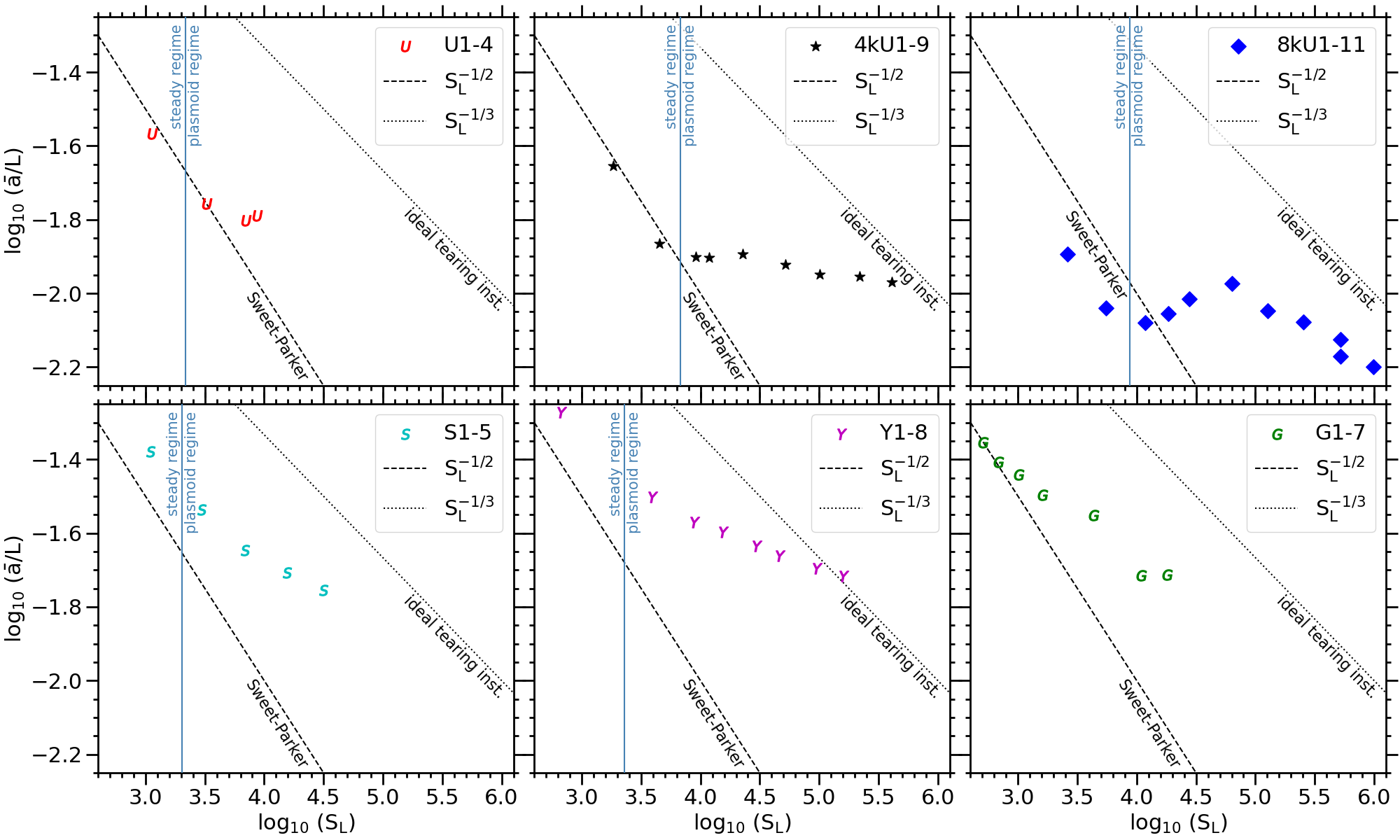}
    \caption{Mean inverse aspect ratio averaged over $t\in[15,35]$ min, plotted against Lundquist number $S_\mathrm{L}$ for each simulation case. The dashed line marks the Sweet-Parker value $a_\mathrm{SP}/L \equiv S_\mathrm{L}^{-1/2}$, and the dotted line shows the ideal tearing instability threshold $a_\mathrm{thr}/L \equiv S_\mathrm{L}^{-1/3}$.}
    \label{fig:aL-S-scatter-plot}
\end{figure*}

In all of our simulation cases, the inverse aspect ratio $\bar{a}/L$ is initially infinitely high, as the current sheet starts at zero length. During the first 15 minutes of the simulation, the aspect ratio decreases rapidly as the current sheet increases in length, reaches an equilibrium value of between 0.005 and 0.05 depending on the simulation case, and remains roughly constant throughout the reconnection phase. For the first 5-10 minutes of each simulation, $\bar{a}/L$ is higher than the ideal tearing instability threshold value $S_\mathrm{L}^{-1/3}$, and the current sheet is stable during this phase (i.e. no plasmoid instability occurs). Shortly after $\bar{a}/L$ passes below $S_\mathrm{L}^{-1/3}$, the current sheet becomes unstable in most of the simulation cases, and plasmoids therefore rapidly appear. However, in a few cases (S1, Y1, U1, 4kU1-2, and 8kU1-2, as discussed below) where the Lundquist number is sufficiently low ($<10^4$), the current sheet remains stable even when $\bar{a}/L < S_\mathrm{L}^{-1/3}$, allowing steady reconnection to occur. Amongst those cases, in the cases with uniform resistivity (U1, 4kU1-2, and 8kU1-2), $\bar{a}/L$ reaches an equilibrium value of close to $S_\mathrm{L}^{-1/2}$, indicating the occurrence of Sweet-Parker reconnection.

In Fig.~\ref{fig:aL-S-scatter-plot},
we show the equilibrium value that $\bar{a}/L$ reaches in each case, which is computed as an average taken over the time interval $t\in[15,35]$ min.
The Sweet-Parker value $a_\mathrm{SP}/L \equiv S_\mathrm{L}^{-1/2}$ is plotted as a dashed line, and the ideal tearing instability threshold $a_\mathrm{thr}/L \equiv  S_\mathrm{L}^{-1/3}$ as a dotted line. All the uniform resistivity cases (top panels) are scattered in a similar manner.
The inverse aspect ratio clearly drops below the ideal tearing instability threshold, allowing plasmoids to appear rapidly in all cases except for those with a sufficiently low Lundquist number to maintain steady reconnection. Those steady-reconnection cases, namely
U1, 4kU1, 4kU2, 8kU1, and 8kU2, all lie just below the Sweet-Parker value in the figure, confirming that these cases indeed follow Sweet-Parker reconnection. 8kU3 also lies just below this line, and U2 on this line, which is in fair agreement with the fact that they lie close to the threshold between the steady regime and the plasmoid-mediated regime. All of the cases that lie within the Sweet-Parker regime are scattered approximately along the Sweet-Parker line, confirming that the inverse aspect ratio is indeed proportional to $S_\mathrm{L}^{-1/2}$ for Sweet-Parker reconnection. 
In the plamsoid-mediated cases (U2-3, 4kU3-9, 8kU4-11), the size of the plasmoids puts a limit on how small the mean thickness $\bar{a}$ of the current sheet can be, and therefore the inverse aspect ratio seems to be almost independent of Lundquist number for those cases. For the nearly shock-mediated cases, 8kU10 and 8kU11, we measured a significantly lower inverse aspect ratio than in the more heavily plasmoid-mediated cases, as the plasmoids here are diminished in size.

As for the anomalous resistivity cases seen in Fig.~\ref{fig:aL-S-scatter-plot}, the inverse aspect ratio decreases slowly with increasing Lundquist number because of a slowly increasing current-sheet length.
In all of the {\syntelis} cases (bottom left panel), the inverse aspect ratio decreases significantly below the ideal tearing instability threshold, which is in close agreement with the fact that plasmoids appear in all cases except S1 (where the sufficiently high resistivity enforces stability of the current sheet). The inverse aspect ratio of S1 is still significantly above the Sweet-Parker value. Therefore, this steady-reconnection case is not Sweet Parker-like, which is expected given that the resistivity is non-uniform.
Among the {\yokshi} cases (bottom centre panel), Y1-5 have an inverse aspect ratio far below $S_{\rm L}^{-1/3}$, and cases Y2-5 are clearly plasmoid-mediaded, as expected, while Y1 has sufficiently high resistivity to maintain steady reconnection, still with $\bar{a}/L > S_{\rm L}^{-1/2}$ (therefore not a Sweet-Parker case). Regarding cases Y6-8, which are also plasmoid-mediated, the inverse aspect ratio drops only barely below $S_{\rm L}^{-1/3}$ in Y6-7 and remains slightly above $S_{\rm L}^{-1/3}$ in Y8. This may indeed explain why the plasmoids in these cases appear diminished in size, indicating a convergence towards shock-mediated reconnection for increasing Lundquist number. 
In all of the {\hdiff} cases (Fig.~\ref{fig:aL-S-scatter-plot}, bottom right panel), $\bar{a}/L$ drops far below the ideal tearing instability threshold, in good agreement with the fact that plasmoids appear relatively large in size in all those cases (as seen in Fig.~\ref{fig:rho_against_length_time}).

\subsection{Reconnection rate}
\label{sec:reconection-rate}

\begin{figure*}[h!]
    \centering
    \includegraphics[width=\textwidth]{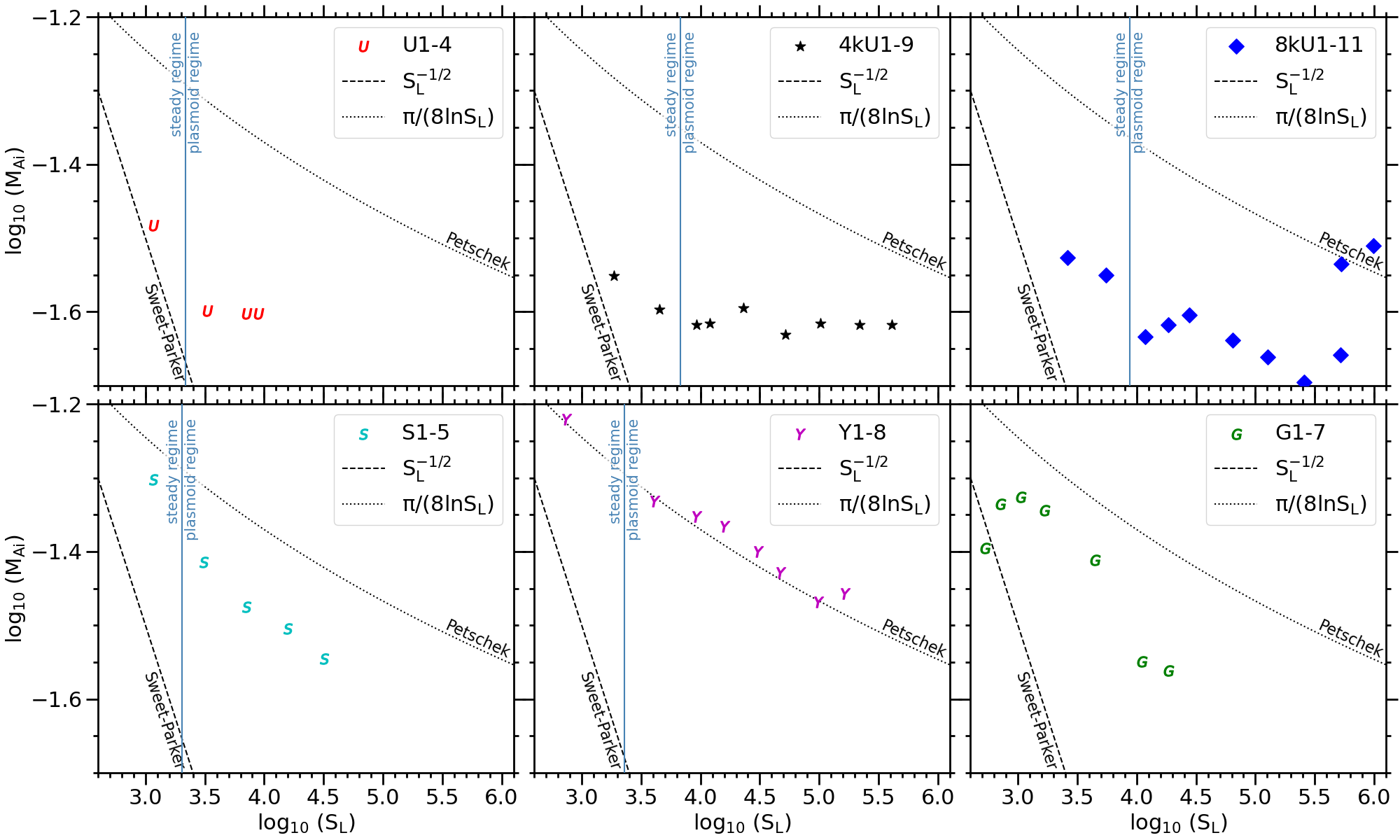}
    \caption{Reconnection rate, averaged over $t\in[15,35]$ min, plotted against Lundquist number $S_\mathrm{L}$ for each simulation case. The dashed line marks the Sweet-Parker value $M_\mathrm{SP} \equiv S_\mathrm{L}^{-1/2}$, and the dotted line the Petschek value $M_\mathrm{Pet} \equiv \pi/8\ln S_\mathrm{L}$.}
    \label{fig:gamma}
\end{figure*}

The reconnection rate $M_\mathrm{Ai} \equiv v_\mathrm{i}/v_\mathrm{Ai}$ of each simulation case is plotted against Lundquist number in Fig.~\ref{fig:gamma}. The Sweet-Parker reconnection rate $M_\mathrm{SP} \equiv S_\mathrm{L}^{-1/2}$ is plotted as a dashed line, and the Petschek reconnection rate $M_\mathrm{Pet} \equiv \pi/8\ln S_\mathrm{L}$ as a dotted curve. Among the uniform resistivity cases (top panels), the reconnection rates of U1, 4kU1, and 8kU1 lie near to the values predicted by the Sweet-Parker model, which is in good agreement with the previously observed Sweet-Parker-like aspect ratio and absence of plasmoids. These cases are therefore indeed in the Sweet-Parker regime. The steady-reconnection cases 4kU2 and 8kU2 are also close enough to the Sweet-Parker line to be characterised as Sweet-Parker reconnection. The plasmoid-mediated cases U2-4, 4kU3-9, and 8kU3-9 lie approximately along the same nearly horizontal line, meaning that the reconnection rate is almost independent of Lundquist number for those cases. A similar change of dependency between reconnection rate and Lundquist number from the Sweet-Parker regime to the plasmoid-mediated regime is seen in the simulations of \citet{2009PhPl...16k2102B}. Cases 8kU10 and 8kU11 both have significantly higher reconnection rates, indeed close to that predicted by the Petschek model, which is in agreement with the fact that these cases are more shock-mediated.
This is due to the fact that the resistivity in these two cases is low enough that the non-uniform viscosity term has a dominating effect on the dynamics of the current sheet. A similar Petschek-like reconnection was seen in the simulations by \cite{2009PhPl...16a2102B}, where a relatively low uniform resistivity was  also applied, and that behaviour was mainly triggered by the non-linear viscosity.

Among the {\syntelis} cases (Fig.~\ref{fig:gamma}, bottom left panel), the steady case S1 has a reconnection rate that is only slightly below the Petschek value, indicating that the reconnection here is nearly Petschek-like. The plasmoid-mediated cases S2 to S5 lie further below the Petschek curve, though the scaling between reconnection rate and Lundquist number is still similar to that of the Petschek model.
Furthermore, all of the {\yokshi} cases (bottom centre panel) lie approximately along the Petschek curve, meaning that their reconnection rates roughly agree with Petschek theory, even though plasmoids are present in all of those cases except for Y1. This agrees perfectly with what \cite{1994ApJ...436L.197Y} found in their 2D simulations of an emerging coronal loop, namely that this anomalous resistivity model is capable of reproducing a non-steady Petschek-like reconnection scheme. Regarding the {\hdiff} cases (bottom right panel), only G1 lies below the Sweet-Parker line. This is in agreement with the fact that the current sheet in this case also has a Sweet-Parker-like aspect ratio, which indicates that non-steady Sweet-Parker reconnection may be occurring here. G3-G5 all have reconnection rates that are slightly below the Petschek value (and G2 somewhere in between), while G6 and G7 have even lower reconnection rates. 

In summary, the reconnection rates obtained with the anomalous resistivity models are in general higher than those obtained with uniform resistivity. The {\yokshi} model is the only one to reproduce reconnection rates that are approximately equal to the Petschek values. The {\hdiff} model, on the other hand, is capable of reproducing relatively high reconnection rates at the same time as reproducing high plasmoid frequencies, as seen in cases G1-5; these latter are the only cases that show reconnection rates above 0.04 whilst also producing more than four plasmoids per minute. 

\subsection{Temperature increase in the reconnection site}
\label{sec:temperature}

\begin{figure}[h!]
    \centering
    \includegraphics[width=\columnwidth]{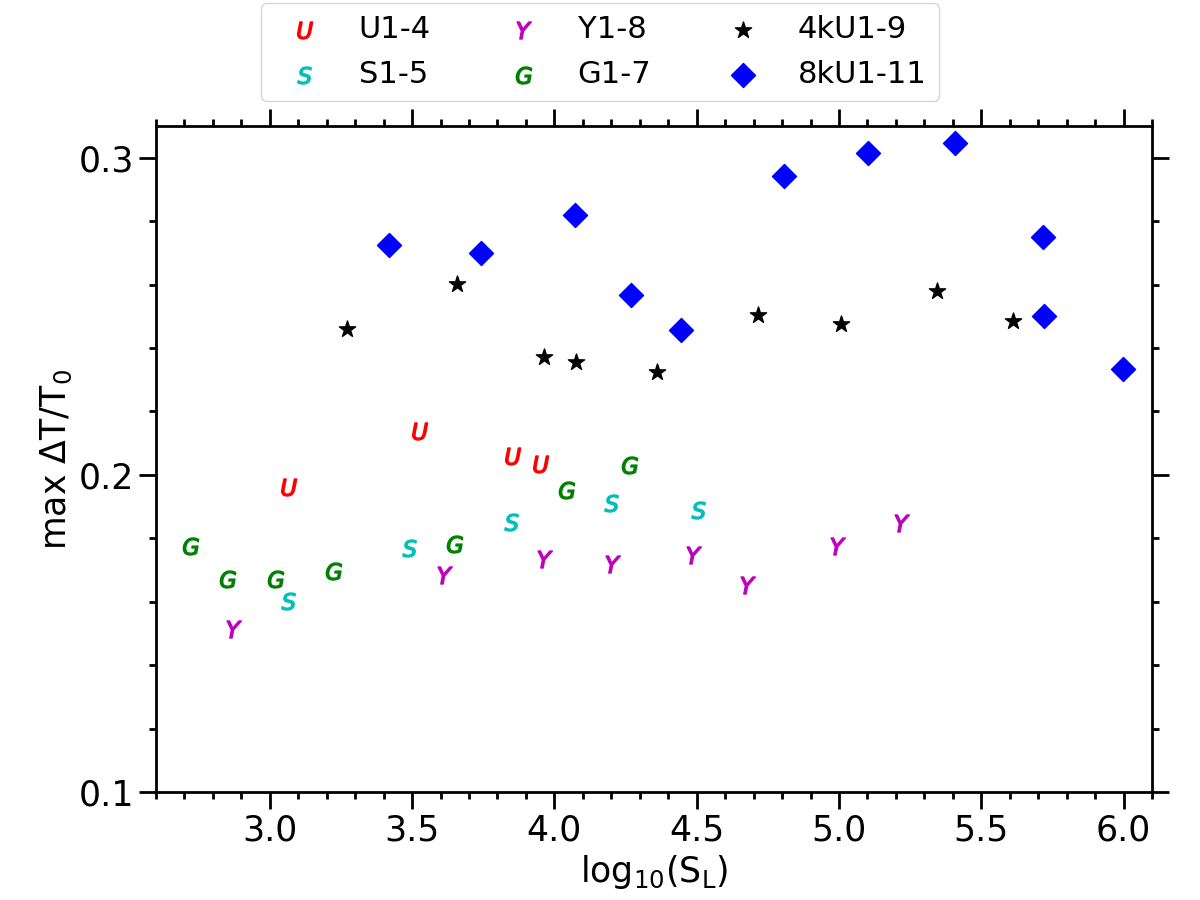}
    \caption{Maximum temperature increase relative to initial temperature, averaged over $t\in[15,35]$ min, plotted against Lundquist number $S_\mathrm{L}$ for all simulation cases.}
    \label{fig:maxtemp}
\end{figure}

As a final step in our analysis of the reconnection process, Fig.~\ref{fig:maxtemp} displays the maximum temperature increase relative to the initial temperature, $\max \Delta T / T_0$, for all simulation cases, which is given by the maximum value of $(T-T_0) / T_0$ found in the computational domain averaged over $t\in[15,35]$ min. This maximum temperature increase lies roughly around 27\%-30\% 
in the 8k cases, at about 25\% in the 4k cases, and between 15\%\ and 22\% in the 2k cases. This shows that the total heating of the current sheet increases with resolution. The reason for this is that the simulation cases with higher resolution obtain significantly longer, though slightly narrower current sheets. Therefore, as the total heating of the current sheet is equal to the heat input per volume integrated over its area, this observed correlation between total heating and resolution is to be expected. Among the uniform resistivity cases, with the exception of 8kU10 and 8kU11, the total heating of the current sheet seems to be almost independent of Lundquist number. This is because the viscous heating of the plasma in the reconnection site, which, predictably, becomes dominant for high Prandtl numbers \citep{2017ApJ...834...10R}, is in our cases found to increase with Lundquist number in a way that balances the corresponding decrease in Joule heating. The nearly shock-mediated cases 8kU10 and 8kU11 have a lower heat input than the other 8k cases because of a significantly shorter and thinner current sheet. 

In all the anomalous resistivity cases, the total heating of the plasma increases weakly with Lundquist number because of the corresponding increase in current-sheet length, as seen in Fig.~\ref{fig:rho_against_length_time}. The scaling between total heating and Lundquist number is strongest in the {\hdiff} cases, and G7 obtain a maximum temperature increase of slightly above 20 \%, reaching the highest temperatures of the anomalous resistivity cases. Among the 2k cases, only the uniform resistivity cases reach higher temperatures, but only at a significantly lower Lundquist number. Therefore, with the resolution of the 2k cases, the {\hdiff} resistivity model is the most suitable for reproducing satisfactorily high temperatures, that is, closer to those obtained in the higher resolution cases, at relatively high Lundquist numbers ($>10^4$).


\section{Discussion}
\label{sec:discussion}

Here, we expand on our previous comparative study of resistivity models  \citepalias{2023A&A...675A..97F} by performing numerical experiments of plasmoid-mediated reconnection in a 2D coronal fan-spine topology. We carried out a parametric study employing the same three anomalous resistivity models as in \citetalias{2023A&A...675A..97F} as well as a model with uniform resistivity. We varied the scaling parameters and the numerical resolution and analysed how the characteristics of the reconnection process depend on Lundquist number.

In all simulations, reconnection occurs along a tilted current sheet in the corona, causing
a temperature increase of 15\%-30 \%. The majority of the experiments show plasmoid-mediated reconnection, regardless of the resistivity model used. Steady reconnection is only found in cases where the resistivity of the current sheet is high enough to prevent plasmoid instability.
The minimum Lundquist number required to reproduce plasmoid instability lies around $2\times 10^3$ in our lowest-resolution cases and converges towards $10^4$ as the resolution reaches sufficiently high values, which is in good agreement with the findings of \citet{2007PhPl...14j0703L}. The hyper-diffusive resistivity model reproduces plasmoid instability at significantly lower Lundquist numbers, which is  due to its dynamic variation in the resistivity along the current sheet.
We also see (in some cases with the drift velocity-dependant resistivity) that the reconnection is shock-mediated rather than plasmoid-mediated if the inverse aspect ratio $\bar{a}/L$ of the current sheet remains above or only slightly below $S_\mathrm{L}^{-1/3}$, indicating that $\bar{a}/L$ has to drop significantly below this threshold in order for the current sheet to become intrinsically unstable, as predicted by \citet{2014ApJ...780L..19P}.

The frequency of plasmoids generated along the current sheet scales with the Lundquist number, following a power law for a certain range of Lundquist numbers. With uniform resistivity, the plasmoid frequency converges towards higher values and a weaker scaling with Lundquist number as the resolution increases. The cases with the highest resolution, $\Delta x = \Delta z = 3.9$ km, reproduce a plasmoid frequency that ranges from 6.9 to 12 plasmoids per minute and scales as $S_\mathrm{L}^{0.210}$ for $S_\mathrm{L} \in [1.8 \times 10^4, 2.6 \times 10^5]$, which is close to the power law found by \citet{2022A&A...666A..28S} for the maximum plasmoid number on a Harris current sheet in a non-adiabatic medium. 
Our simulated plasma is also non-adiabatic, which explains why we reproduce a scaling law  here that is similar to theirs rather than to those derived in the adiabatic cases of \citet{2007PhPl...14j0703L} and \citet{2010PhPl...17f2104H}, where the plasmoid number was $\propto S_{\rm L}^{0.375}$ in the linear reconnection phase and $\propto S_{\rm L}$ in the non-linear phase.
For $S_\mathrm{L} < 10^4$, steady Sweet-Parker reconnection occurs that is characterised by the absence of plasmoids, a Sweet-Parker-like aspect ratio of the current sheet, and a reconnection rate similar to that predicted by the Sweet-Parker model. For sufficiently high Lundquist numbers ($S_\mathrm{L} > 5\times 10^5$), a rather shock-mediated Petschek reconnection occurs, which is similar to what was found by \citet{2009PhPl...16a2102B}, with a nearly adiabatic scaling between shock frequency and Lundquist number and a reconnection rate close to the Petschek value.
This happens because the resistivity here  is low enough to allow the non-uniform viscous term to dominate.

Among our simulation cases with the lowest resolution, $\Delta x = \Delta z = 15.6$ km, plasmoid-mediated reconnection is  reproduced for only a narrow range of Lundquist numbers ($S_L\in [3\times 10^3, 10^4]$) with uniform resistivity. The anomalous resistivity models help to increase this range. The drift-velocity-scaled model (\yokshi) used by \citet{1994ApJ...436L.197Y} reproduces Petschek reconnection for any Lundquist number (being steady for $S_\mathrm{L} < 10^3$) with reconnection rates approximately equal to $\pi/(8\ln S_\mathrm{L})$ and a nearly adiabatic scaling between plasmoid (or shock) frequency and Lundquist number. The model with 
resistivity proportional to current density (\syntelis) reproduces similar results, but on a narrower range of Lundquist numbers, with a lower plasmoid frequency that scales more closely with Lundquist number and a reconnection rate that is slightly lower than the Petschek value. The hyper-diffusive resistivity model of Bifrost (\hdiff) reproduces higher plasmoid frequencies (4.2-7.6 plasmoids per minute) with a weaker scaling with Lundquist number ($\propto S_\mathrm{L}^{0.142}$) than any of the other resistivity models applied on the same resolution; indeed, it is the only resistivity model that, for the given resolution, reproduces a plasmoid frequency with a weaker scaling to Lundquist number than the $S_\mathrm{L}^{0.375}$ scaling predicted for adiabatic reconnection  \citep{2007PhPl...14j0703L}. This resistivity model therefore reproduces plasmoid characteristics that more closely resemble those seen in the higher-resolution cases. It is also the only resistivity model that reproduces both relatively high reconnection rates ($> 0.04$) and plasmoid frequencies (> 4 plasmoids per minute) at the same time. Additionally, for significantly high Lundquist numbers (>$10^4$), the hyper-diffusive
resistivity model of Bifrost reproduces a higher total heating of the plasma than the other resistivity models applied on the same resolution, reaching temperatures closer to those of the higher-resolution cases.  Therefore, this model indeed  proves to be suitable for simulating dynamic plasmoid-mediated reconnection, and is also applicable for 3D models of the solar atmosphere without requiring extremely high resolution. Indeed, this model has been successfully used for simulations of flux emergence with plasmoid reconnection leading to EBs and UV bursts \citep{2019A&A...626A..33H} as well as nanoflare-like events with synthesised line spectra detectable for the upcoming MUSE mission \citep{2023A&A...677A..36R}.

The most important result of this comparative study is that, out of the four resistivity models applied on the same reconnection experiment with the same numerical resolution, the plasmoid characteristics produced with the hyper-diffusive model most closely resemble those obtained with uniform resistivity with significantly higher resolution. Additionally, by taking into account scaling laws previously derived for spontaneous reconnection on Harris sheets \citep{2007PhPl...14j0703L, 2009PhPl...16k2102B, 2010PhPl...17f2104H, 2022A&A...666A..28S}, we show that we are able to derive very similar scaling laws for a more driven reconnection process. This indicates that such scaling laws may apply on a wider range of reconnection processes, allowing us to better understand more complex scenarios such as reconnection driven by granular motion \citep{2022ApJ...935L..21N}.

The complex behaviours of plasmoid instability may only be fully understood through three-dimensional numerical studies; namely the turbulent splitting, kinking, and merging of plasmoids seen in the coronal mass ejection simulation of \citet{2023ApJ...955...88Y}, or the chaotic tearing-thermal instability leading to coronal condensation similar to prominences and coronal rain blobs simulated by \citet{2023A&A...678A.132S}. Two-dimensional particle-in-cell (PIC) simulations of waves in plasmoid-mediated reconnection have provided new insights into the different natures of waves inside and outside current sheets as an effect of the tearing instability \citep{2022FrASS...8..237S}. High-resolution 2D MHD simulations with resistivity predicted from particle-collision probabilities including radiative cooling and partially ionised effects have provided detailed information on the energy balance in plasmoid reconnection in the chromosphere leading to EBs \citep{2023RAA....23c5006L} and UV bursts \citep{2022A&A...665A.116N}. Though MHD simulations with anomalous resistivity may lead to a slightly more approximate representation of the reconnection process, this study proves that the hyper-diffusion model of Bifrost is indeed helpful in numerically studying phenomena on the Sun that would otherwise require a significantly higher resolution to simulate with a low, Spitzer-like resistivity.

%
%
\begin{acknowledgements}
This research has been supported by the European Research Council through the
Synergy Grant number 810218 (``The Whole Sun'', ERC-2018-SyG) and 
by the Research Council of Norway through its Centres of Excellence scheme, project
number 262622.
The simulations were performed on resources provided by  Sigma2 - the National Infrastructure for High Performance Computing and Data Storage in Norway.
The authors are grateful to the referee for his/her constructive comments to improve the paper.
\end{acknowledgements}

\bibliographystyle{aa}
\bibliography{Faerder2023_paper2}

\end{document}